\documentclass[cits]{PoS}
\usepackage{amssymb}
\usepackage{amsmath}
\usepackage{booktabs}

\title{
\begin{picture}(0,0)(0,0)%
   \put(350,75){\makebox(0,0)[l]{\textnormal{\normalsize KEK-CP-309}}}%
   \put(350,60){\makebox(0,0)[l]{\textnormal{\normalsize OU-HET-830}}}%
   \end{picture}%
Axial U(1) symmetry at finite temperature with M\"{o}bius domain-wall fermions}

\ShortTitle{Axial U(1) symmetry at finite temperature}

\author{JLQCD Collaboration: \speaker{Guido Cossu}$^a$\thanks{E-mail: cossu@post.kek.jp}, Hidenori Fukaya$^c$, Shoji Hashimoto$^{a, b}$, Takashi Kaneko$^{a, b}$, Jun-ichi Noaki$^a$, Akio Tomiya$^c$\\
\\
\llap{$^a$}Theory Center, IPNS, High Energy Accelerator Research Organization
  (KEK), Tsukuba 305-0801, Japan\\
\llap{$^b$}School of High Energy Accelerator Science, The Graduate University for Advanced Studies (Sokendai), Tsukuba 305-0801, Japan\\
\llap{$^c$}Department of Physics, Osaka University, Toyonaka 560-0043, Japan}

\abstract{We investigate the axial $U(1)$ symmetry restoration at finite temperature in two flavor QCD. We employ the M\"{o}bius domain-wall formalism that is designed to achieve good chiral symmetry. We show the measurements of a difference of meson susceptibilities, sensitive to the \ua symmetry breaking. The signal is dominated by zero and near-zero modes. By reweighting the measure to that of overlap fermions we find a suppression of the \ua breaking effects above the chiral transition temperature. }

\FullConference{The 32nd International Symposium on Lattice Field Theory\\
		 23-28 June, 2014\\
		 Columbia University New York, NY}

\newcommand{\ua}{$U(1)_A$~}

\begin{document}
\section{Introduction}

The question whether the \ua symmetry is effectively restored above the chiral phase transition is still open. In the well-known pattern of symmetry breaking in $N_f$ flavor QCD at low temperature 
\begin{equation}
SU(N_f)_L \otimes SU(N_f)_R \otimes U(1)_V \otimes U(1)_A \rightarrow U(1)_V \otimes SU(N_f)_V ,
\end{equation}
the \ua symmetry is peculiar since it is violated by the quantum anomaly. It is coming from the presence of gauge configurations with non-zero topological charge $Q$ that generate an anomalous contribution to the divergence of flavor-singlet axial-vector current \cite{'tHooft:1976up}.

The answer to this question may have an impact also from a  phenomenological viewpoint: the order and the universality class of the phase transition depend on whether the axial symmetry is restored or not \cite{Pisarski:1983ms,Pelissetto:2013hqa}.
Models like the instanton gas \cite{Gross:1980br} predict a suppression of the instanton density and thus an effective restoration of the \ua symmetry at very high temperatures $T \gg T_c$\footnote{$T_c$ is the temperature of the chiral phase transition, namely the location of the peak of the susceptibility of the chiral condensate.}, in the domain of their applicability. Only recently the lattice QCD studies on this subject have been (re)started at around the phase transition using several formulations for the fermion action and focusing on  different observables \cite{Cossu:2013uua,Buchoff:2013nra,Brandt:2013mba,Sharma:2013nva,Chiu:2013wwa}.

In the previous JLQCD work we studied the problem using the overlap fermion formulation \cite{Cossu:2013uua}. This guarantees exact chiral symmetry of the lattice action in the chiral limit. The Dirac spectrum and the meson correlator measurements both indicate a restoration of the \ua symmetry in QCD with two degenerate flavors. A gap in the spectrum opens at temperatures above $T_c$ when the quark mass is decreased toward the chiral limit. At the same time, the disconnected diagrams vanish, leading to a degeneracy of the correlators of the lightest mesons, which is a signal of the restoring symmetry. The problem was also studied theoretically in \cite{Aoki:2012yj} showing that with two degenerate flavors the spectral density of the Dirac operator behaves like $\rho(\lambda) \sim c\lambda^3$ in the high temperature phase. It implies that the \ua anomaly is invisible in the meson susceptibilities. This result is compatible with our lattice simulations.  

The most important source of systematic errors in the previous project was the need to fix the global topology $Q$. In order to avoid this limitation we started a new series of simulations using the M\"{o}bius domain-wall fermion formulation \cite{Kaneko:2013jla} with our new code platform IroIro++ \cite{Cossu:2013ola}. Compared to the standard domain-wall formulation we have the advantage of having smaller residual mass, i.e. better chiral symmetry. As we are showing in these proceedings, a precise chiral symmetry is quite relevant for the study of the \ua problem and even M\"{o}bius fermions would not be sufficient. Another important issue is the mass dependence: we only observe the restoration when approaching the chiral limit. The current results are in accordance with the outcome of the previous overlap project.

In the following sections we present the methodology of our analysis and discuss the results to date. Then we draw some conclusions based on the current data and a perspective on the ongoing simulations.

\section{Analysis}

We are studying $N_f=2$ QCD with tree-level Symanzik improved gauge action and smeared M\"{o}bius domain-wall fermions. The details of this fermonic action were reported in the proceedings of the previous lattice conference \cite{Kaneko:2013jla} and are the same as our zero temperature simulations \cite{Noaki:2014ura}. Simulation points cover a region of temperatures between 150 and 250 MeV with up to three different masses for the points just above the phase transition. The measured residual mass above the phase transition is less than 1 MeV for the $N_t=8$ runs~\cite{HashimotoResMass}.

We measure two main observables related to the axial $U(1)$ symmetry: the eigenvalue spectrum $\rho(\lambda, m)$ of the hermitian Dirac operator ($H \equiv \gamma_5 D$) and the \ua susceptibility $\Delta$ defined as a difference of the susceptibilities of $\pi$ and $\delta$ channels

\begin{equation}
\Delta = \chi_\pi - \chi_\delta = \int {\rm d}^4x \langle \pi^a(x) \pi^a(0) - \delta^a(x)\delta^a(0) \rangle.
\label{eq:Delta}
\end{equation}
It vanishes when the \ua symmetry is fully restored in the vacuum. This quantity has a simple representation in terms of the Dirac operator eigenvalue spectrum:

\begin{equation}
\Delta = \lim_{m \rightarrow 0}\lim_{V\rightarrow \infty} \int \frac{2m^2 \rho(\lambda,m)}{(\lambda^2+m^2)^2} {\rm d}\lambda = \lim_{m \rightarrow 0}\lim_{V\rightarrow \infty} \Bigl(\frac{2N_0}{Vm^2} + \sum_{\lambda_i \neq 0} \frac{2m^2(1-\lambda_i^2)^2}{V(\lambda_i^2 + m^2)^2(1-m^2)^2} \Bigr )
\label{eq:DeltaSpectr}
\end{equation}
where the limits must be taken in that order and $N_0 = |Q|$, the number of zero eigenvalues of $D$. The second equation comes from an expansion in the eigenvalues $\lambda_i$ of the discretized overlap operator.

We concentrate on $\Delta$ and the discussion on the spectrum of the Dirac operator is given in another paper of these proceedings \cite{Akio}.

\subsection{Current results}

The measurement of $\Delta$ is quite delicate and the details of the method could affect the final result. We observe that a simple integration of the correlator from a local source is highly sensitive to the position of the source. This is explained by the spatial location of the zero and lowest-lying modes of $H$. A source hitting one of these modes would overestimate the final result and viceversa if far away. A stochastic estimate using a $Z_2$ noise source all over the volume is more stable.

\begin{figure}[t]
  \centering
  \includegraphics[clip=true, width=0.49\columnwidth]{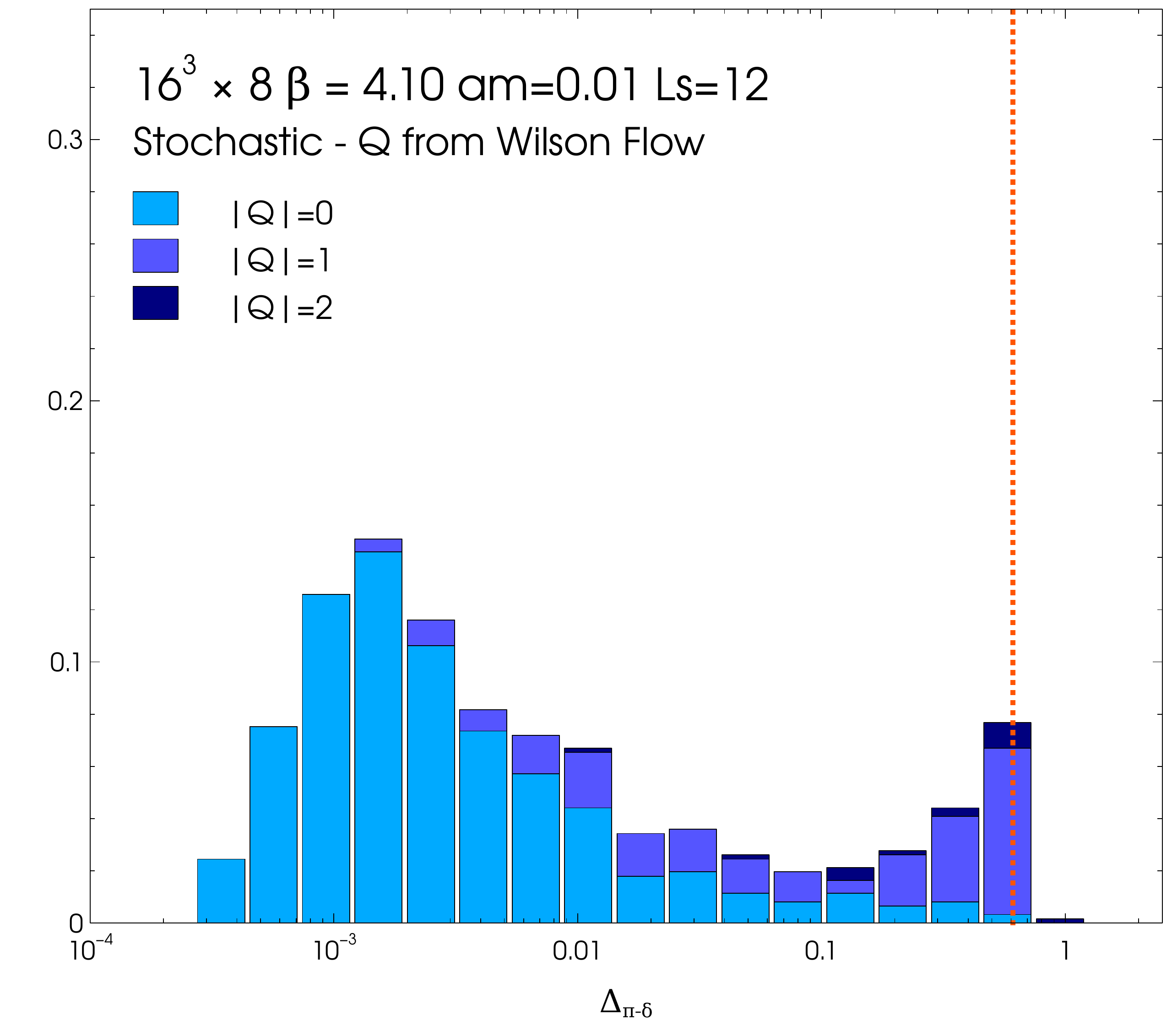}
  \includegraphics[clip=true, width=0.49\columnwidth]{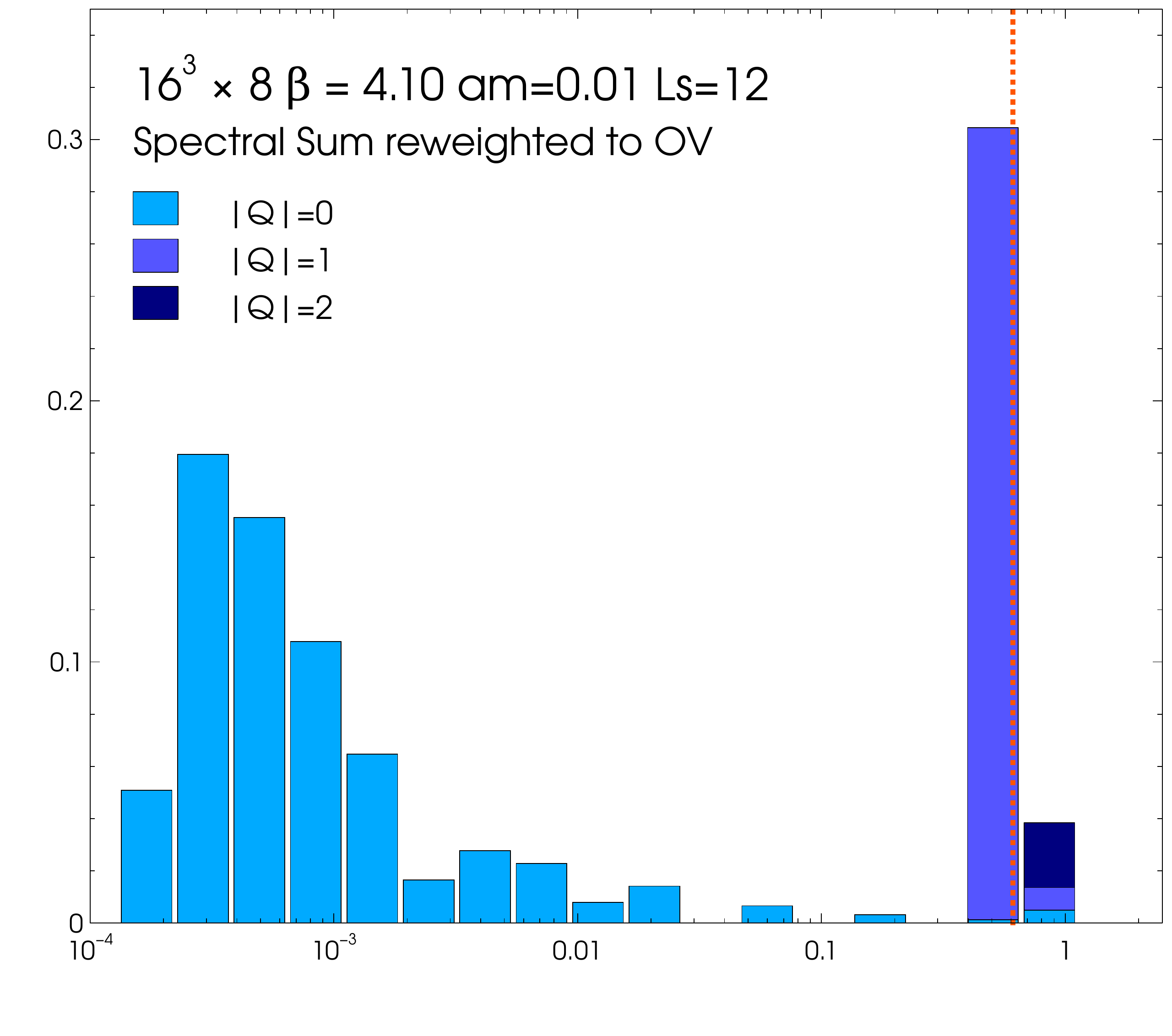}
  \caption{Histogram of $\Delta$ on a $16^3\times 8$ lattice at $\beta=4.10$, $T \simeq 200$ MeV (left). Horizontal axis is in the log-scale. The colors identify the topological sectors $Q$ calculated after the Wilson flow. Dashed vertical line is at $\Delta = 2/(Vm^2)$. The histogram after reweighting the spectrum to the overlap formalism (right).}
  \label{fig:Spect_vs_TopWF} 
\end{figure}

Before taking the limits in (2.2), there are large contributions in multiples of $2/(Vm^2)$, depending on the number of zero modes, when $m$ is smaller than the smallest non-zero eigenvalues. Such correlation with the topological charge is not well respected in the data as shown in the left panel of Figure~\ref{fig:Spect_vs_TopWF}. The plot shows the histogram of $\Delta$ as calculated with the stochastic method. The $\Delta$ in the horizontal axis is in a logarithmic scale. Contributions from different topological sectors $Q=0,1,2$ are shown by different colors. The measurement of $\Delta$ yields results ranging over three orders of magnitude. Among those, the contribution from configurations with $Q\neq 0$ should fall above the dashed thick line that represents the location of the expected contribution from one zero mode, i.e. $2/(Vm^2)$.  Violations of the Ginsparg-Wilson relation and imperfect estimate of the topological charge\footnote{The topological charge $Q$ is estimated after some steps of Wilson flow using the simplest lattice discretization of $F \tilde F$.} are the source of this inconsistency. It already suggests that the final result for this quantity is strongly affected by how well the chiral symmetry is realised. 

A preliminary outcome of this analysis is that the contribution from configurations with $Q \neq 0$ is dominant, 76\% of the signal for the case shown in Figure~\ref{fig:Spect_vs_TopWF}. Nevertheless, we expect from~(\ref{eq:DeltaSpectr}) that this contribution vanishes as $1/\sqrt V$ in the large volume limit since the topological susceptibility, $\sim N_0^2/V$, stays constant in that limit. 

\begin{figure}[t]
  \centering
  \includegraphics[clip=true, width=0.48\columnwidth]{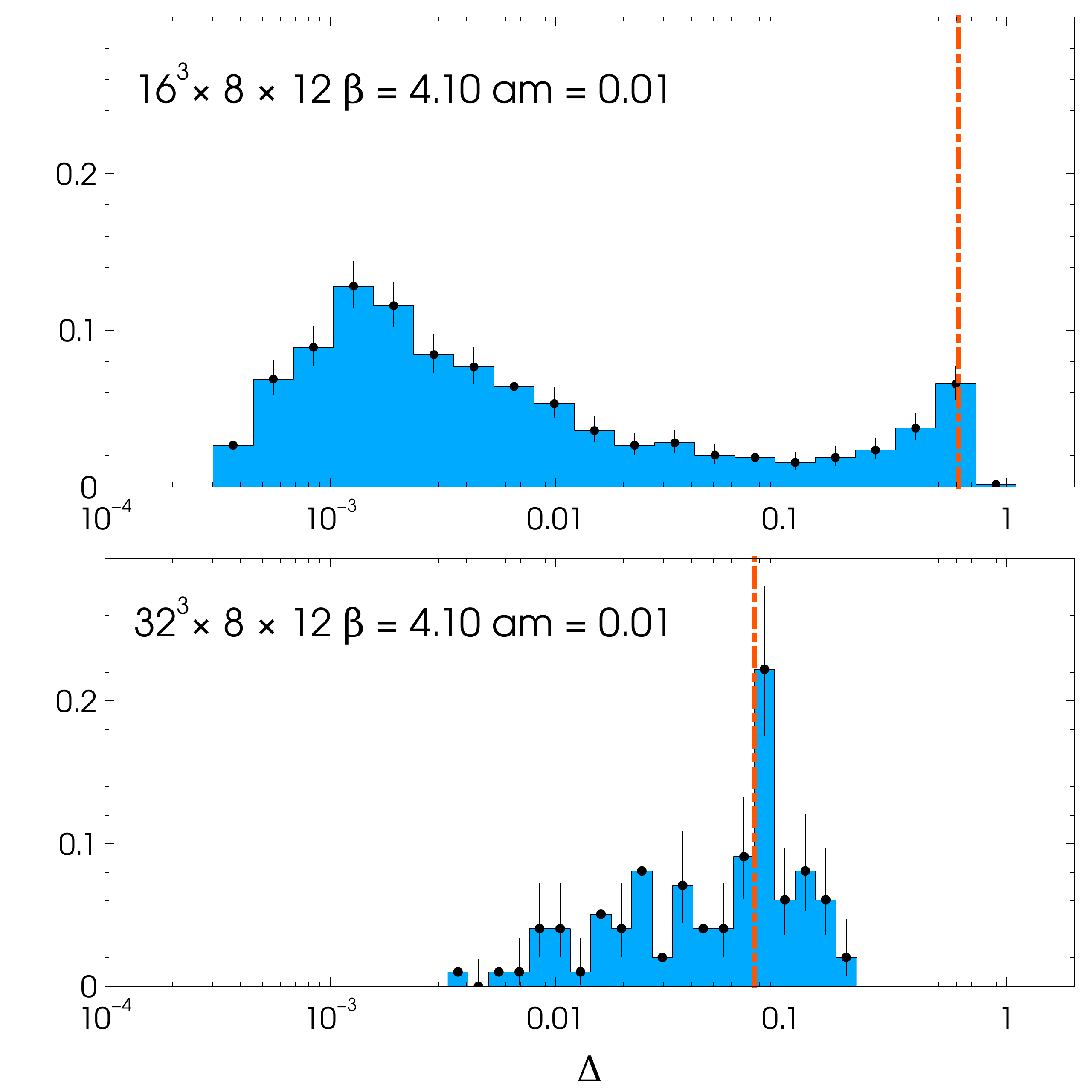}
  \caption{Histograms at $\beta=4.10$, $T=200$ MeV. Volume dependence of the stochastic estimate of $\Delta$. Upper panel shows the data on the $16^3\times 8$ lattice, while the bottom panel if for $32^3\times 8$. The dashed vertical lines show the location of $\Delta = 2/(Vm^2)$. }
  \label{fig:Spect_VolumeDep} 
\end{figure}

Figure~\ref{fig:Spect_VolumeDep} shows the volume dependence of the $\Delta$ susceptibility. The position of the peak due to zero modes as expected from~(\ref{eq:DeltaSpectr}), $2/Vm^2$, is shown by dashed lines. It is evident that the bulk part of the signal (below the zero mode contribution) is increasing with the volume, suggesting that the dominance of the zero mode is indeed an artifact of the finite volume. 

\begin{figure}[th]
  \centering
  \includegraphics[clip=true, width=0.49\columnwidth]{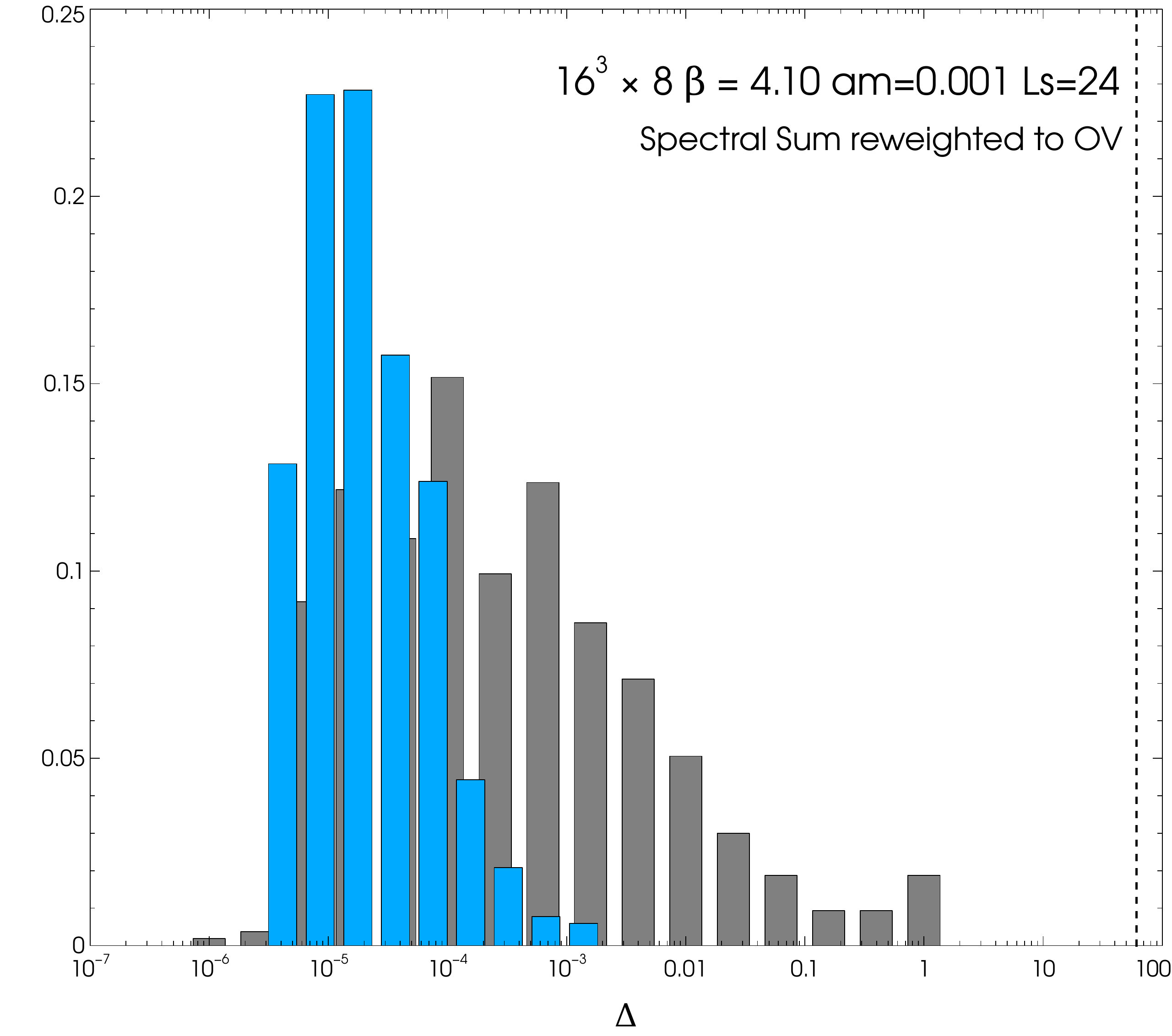}
  \caption{Histogram of $\Delta$ without (grey) and with (blue) the overlap reweighting on the $16^3\times 8$ lattice at $am = 0.001$. Horizontal axis in log-scale. Histograms of the spectral sum estimate of the $\Delta$ defined in~(\protect\ref{eq:DeltaSpectr}). Dashed vertical line is at $\Delta = 2/(Vm^2)$. Only the $Q=0$ configuration are present in the reweighted histogram as a result of the reweighting procedure.}
  \label{fig:Spect_vs_TopWF_0.001} 
\end{figure}

Since most of the signal originates from the zero modes and the lowest part of the spectrum, we examined the possibility that they are affected by the violations of the Ginsparg-Wilson (GW) relation. Evaluating the GW relation on the eigenmodes of $H$ we estimate the size of the violation for each mode and find that the lowest part of the spectrum shows more significant violation compared to the average eigenmode (see \cite{Akio} for a detailed discussion). Motivated by these results we attempt the reweighting of the results from our action to the other one that satisfies the GW relation exactly. We call it the overlap action below, but it has the same kernel as the domain-wall fermion. Only the sign function approximation is improved by treating the low-lying eigenvalues of the kernel exactly. We calculate the reweighting factor as described in \cite{Fukaya:2013vka}. The reweighted histogram of the spectral sum representation of $\Delta$ is shown in Figure~\ref{fig:Spect_vs_TopWF}, right panel. The zero topology contribution to the signal is pushed down toward the left of the plot by the reweighting. This is the part that is expected to survive in the thermodinamical limit. On the other hand, the contribution from the non-zero topological charge concentrates on the bins of $2N_0/Vm^2$ as expected. A more striking result is obtained when the mass is decreased to $am=0.001$, Figure~\ref{fig:Spect_vs_TopWF_0.001}. By the reweighting the signal is suppressed by about three orders of magnitude. We obtain similar results for temperatures close to the phase transition. These results agree with the previous conclusions with overlap fermions~\cite{Cossu:2013uua} and indicate a strong suppression of $\Delta$ in the chiral limit. The full analysis for this observable on a $32^3 \times 8$ lattice to check the finite volume effects is ongoing.

\subsection{Profile of the near-zero modes}

We have seen that the lowest modes dominate the violation of the \ua symmetry, as in~(\ref{eq:DeltaSpectr}). The following question now concerns the nature of these near zero modes. We study some basic quantities such as the participation ratio $(V \sum \psi(x)^4)^{-1}$ and the unfolded level density to probe the geometrical properties of these modes. From the scaling of the participation ratio with the volume we find that the lowest modes in the spectrum are localized and Poisson distributed (also found in \cite{Giordano:2013taa}). This suggests that they are independent from each other and having zero or minimal interactions. The higher part of the spectrum is instead following the predictions from a random matrix model, i.e. the gaussian unitary ensemble.

\begin{figure}[t]
  \centering
  \includegraphics[clip=true, width=0.49\columnwidth]{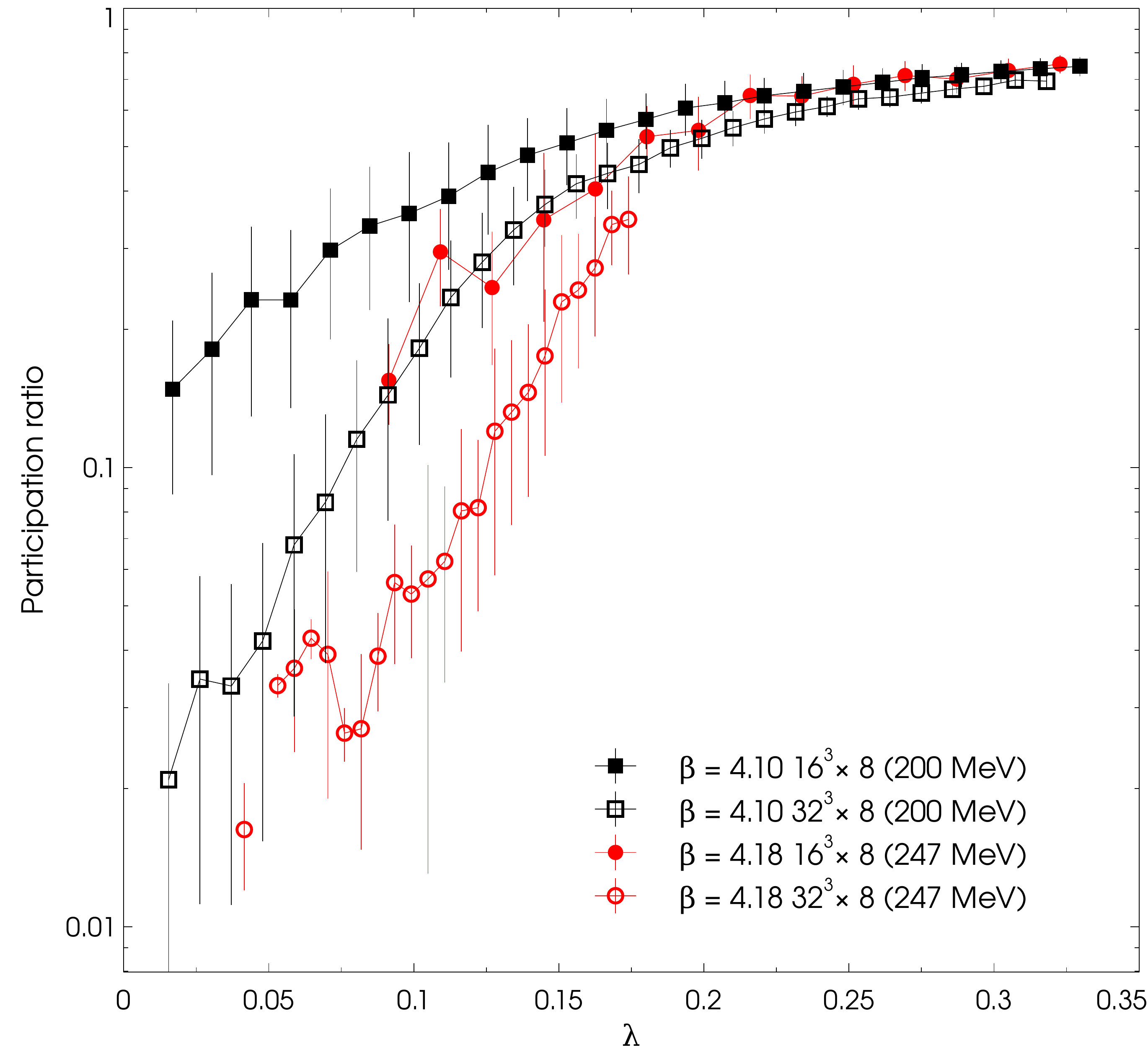}
  \includegraphics[clip=true, width=0.49\columnwidth]{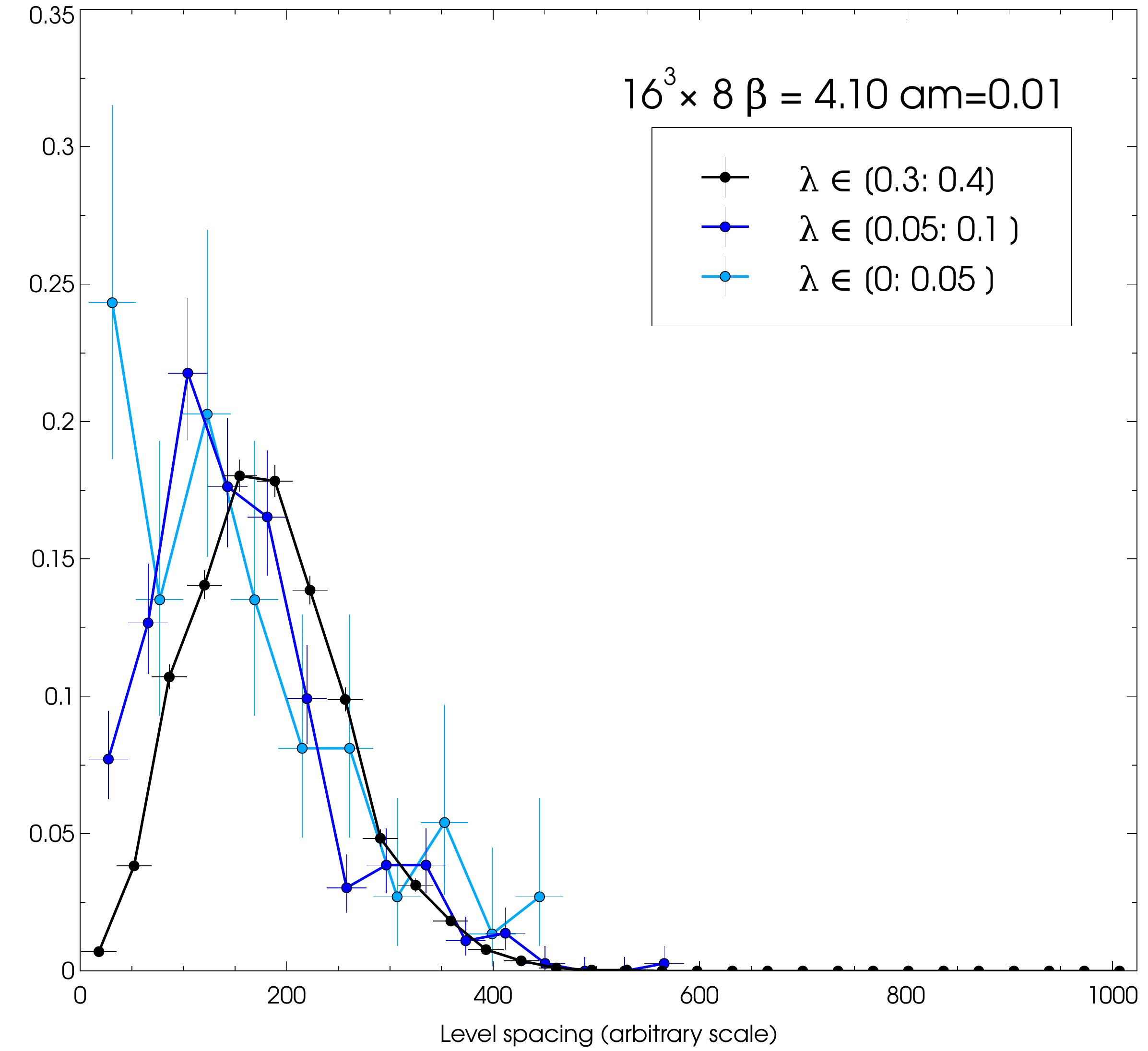}
  \caption{(left) Participation ratio for two temperature and lattice sizes. Increasing the lattice size the relative volume occupied by the lowest modes, $\lambda \lesssim 0.05$ for $\beta=4.10$ and $\lesssim 0.1$ for $\beta=4.18$, decreases as $1/V$, showing that the modes have a definite physical size. Higher modes show the typical behavior of extended modes, by not showing any volume scaling. (right) Unfolded level spacing distribution for 3 regions in the Dirac spectrum. The lowest region exhibits a Poisson-like distribution and the higher modes follow the gaussian unitary ensemble prediction.}
  \label{fig:IPR} 
\end{figure}

The spatial distribution of the norm $\psi(x)^\dagger \psi(x)$ and the pseudoscalar density $\psi(x)^\dagger \gamma_5 \psi(x)$ operator for these low modes shows that they are composed by two lumps of different chirality very close to each other. This is a preliminary observation that suggests objects with definite chirality and small, or absent, interaction. They look similar to an instanton-antiinstanton pair but also a dyonic source is not excluded.

\section{Conclusions and perspective}

Using M\"{o}bius domain wall fermions we have shown an evidence that the \ua symmetry breaking is suppressed in the high temperature phase. The difference of the susceptibilities of the $\pi$ and $\delta$ channels decreases as we approach the chiral limit after reweighting to the overlap fermion action. The zero-mode contribution obeys the theoretical expectation and vanishes in the thermodynamical limit. Breaking of the \ua symmetry originates from the lowest part of the spectrum and these modes are highly sensitive to the violation of the chiral symmetry, even with the M\"{o}bius domain-wall fermion action that is designed to reduce such effects. The reweighting eliminates the effect of such artefacts and as a result the \ua breaking is suppressed.
These results are mirrored by the Dirac operator spectrum calculation that is presented in~\cite{Akio}.
We are currently running the analysis on a bigger volume ($32^3\times 8$) and on a finer lattice ($32^3\times 12$) to confirm these observations.

\acknowledgments
Numerical simulations are performed on IBM System Blue Gene Solution at KEK. We thank P. Boyle for helping in the optimization of the code for BGQ. This work is part of the Large Scale Simulation Program 12/13-04 and 13/14-04 and is supported in part by the Grand-in-Aid of the Japanese Ministry of Education (No. 26247043, 26400259, 25800147) and  SPIRE (Strategic Program for Innovative Research) Field 5.

\bibliographystyle{hunsrt}
\bibliography{proceedingsU1.bib}

\end{document}